\documentclass[12pt,fleqn]{article}
\usepackage{latexsym}
\newtheorem{thm}{Theorem}[section]
\newtheorem{prop}[thm]{Proposition}
\newtheorem{lemma}[thm]{Lemma}

\newcommand{\semi}{\mbox{$\times\!\rule{0.3pt}{1.1ex}\:$}}
\def\la{\langle}
\def\ra{\rangle}
\def\i{^{-1}}
\def\p{^\phi}
\def\ie{\emph{i.e.}}
\def\eq{\begin{equation}}
\def\en{\end{equation}}
\def\ot{\otimes}
\def\id{\mbox{\small id}}
\font\gt = eufm10 scaled \magstep 1
\def\gtg{\mbox{\gt g}}
\def\gtgr{\mbox{{\gt g}\hspace{.5pt}$_r$}}
\begin{document}
\begin{titlepage}
\rightline{LMU-TPW 97-1}
\rightline{CRM-2448}
\rightline{solv-int/9701011}
\vspace{2em}
\begin{center}{\bf\large Twisted Quantum Lax Equations}\\[2em]
Branislav Jur\v co${}^{*}$ and Peter Schupp${}^{**}$\\[2em]
{\sl ${}^*$CRM, Universit\'e de Montr\'eal\\Montr\'eal (Qc), H3C 3J7, Canada}\\[2em]
{\sl ${}^{**}$Sektion Physik der Universit\"at M\"unchen\\
Theresienstr. 37, 80333 M\"unchen, Germany} \\[6em]
\end{center}
\begin{abstract}
We give the construction of twisted quantum Lax equations associated with
quantum groups. We solve these equations using factorization properties
of the corresponding quantum groups. Our construction generalizes in many 
respects the AKS construction for Lie groups and the construction of
M.\ A.\ Semenov Tian-Shansky for the Lie-Poisson case.
\end{abstract}

\vfill
\noindent \hrule
\vskip.2cm
\hbox{{\small{\it e-mail: }}{\small \quad jurco@CRM.UMontreal.CA,\quad
peter.schupp@physik.uni-muenchen.de}}
\end{titlepage}
\newpage
\setcounter{page}{1}
\section{Introduction}

In this paper we discuss the quantum integrable systems related
to quasitriangular Hopf algebras in a way that generalizes the classical 
theory based on the construction of Adler \cite{Ad}, 
Kostant \cite{K}, Symes \cite{Symes} for the 
Lie groups and its subsequent generalization to the 
Lie-Poisson case due to Semenov-Tian-Shansky \cite{STSDress}. 
The classical theory which we briefly summarize in this section 
gives the construction and solution
of integrable systems possessing a (twisted) Lax pair and $r$-matrix
formulation. The rich structure of these integrable systems appears 
naturally as a consequence of the factorization
properties of the groups under consideration. 
Within this approach the fundamental methods (inverse scattering  
method \cite{Miura}, algebro-geometric methods of solution) 
and the fundamental  notions of the soliton theory, such as 
$\tau$-function \cite{JM} and  
Baker-Akhieser function \cite{DKN}, \cite{SW}, found their unifying and 
natural group-theoretical explanation.
This paper deals with the quantum case. 
The theory of integrable models in quantum mechanics and quantum field theory 
made remarkable progress with the quantum version of  
the inverse scattering method, which goes back to the seminal Bethe  
ansatz for the solution of the Heisenberg
spin chain. We refer the reader for a review of related topics to the  
books \cite{Gaudin}, \cite{BIK} and to the papers \cite{F}, \cite{KS},
\cite{Thacker}. This development suggested the introduction of 
quantum groups \cite{DrQG}, \cite{J} as  
algebraic objects that play in the quantum case a role analogous to that of  
Lie groups in the classical theory. However, we were still  
missing (with the exception of the quantum integrable systems with  
discrete time evolution \cite{Resh}) a quantum analogue of the  
factorization theorem for the solution of the Heisenberg equations of  
motion for quantum integrable systems. We have also to mention the 
remarkable paper \cite{Maillet} in this context.

The quantum systems we consider are quantum counterparts of 
those described by the classical factorization theorem. 
In a recent paper by M. Schlieker and one of the coauthors \cite{JSch2} 
the quantum version of the theory in the case without twisting
was formulated.
The main result of the present article, 
the quantum factorization theorem (formulated in the Section 4),
as well as the remaining discussion extends all constructions in the presence
of twisting, which seems to be nontrivial.

\subsection{Classical integrable systems}

Here we briefly review the construction of integrable systems and
their solution by factorization which is due to
M. Semenov-Tian-Shansky \cite{STSDress} and
which generalizes to the case of the Poisson Lie groups
the construction of Adler, Kostant and Symes \cite{Ad}, \cite{K}, \cite{Symes}.
Let $G$ be a quasitriangular Poisson Lie group, which is for simplicity
assumed to be a matrix group. Let \gtg\ 
be the corresponding Lie algebra and
$r\in \gtg \otimes \gtg$ the classical $r$-matrix, a
solution to the classical Yang-Baxter equation.
In the following we will use a notation that does
not distinguish between
the universal element and its matrix representative.
We will denote by $G_r$ and $\gtgr$ 
the dual Poisson Lie group and its 
Lie algebra respectively. The pairing between \gtg\ and \gtgr\
is denoted by $\langle.,.\rangle$.
The Poisson structure on $G$ is given by the Sklyanin bracket
\begin{equation}
\{g_1,g_2\}=[r, g\otimes g]\,.\label{Sklyanin}
\end{equation}
Here we used the standard tensor notation:
$g_1=g\otimes 1$, $g_2=1\otimes g$, with
$g\in G$ being a group element (matrix) 
and $1$ the unit matrix. The commutator
on the right-hand side is the usual matrix 
commutator in $\gtg \otimes \gtg$.
With the universal $r$-matrix $r$ we can 
associate two mappings $r_{\pm}$:
$\gtgr \rightarrow \gtg$
\eq
r_+(X) = \langle X\otimes id, r\rangle\equiv X_+\,,\qquad
r_-(X) = -\langle id \otimes X, 
r\rangle\equiv X_- \label{maps}\,.
\en
These mappings are algebra homomorphism.
Let $\gtg\hspace{.5pt}_{\pm}=\mbox{Im}(r_{\pm})$
be the corresponding subalgebras in \gtg. 
Consider the combined mapping
$i_r = r_- \oplus r_+:\, \gtgr \rightarrow \gtg \oplus \gtg$.
Let us assume that the mapping
$r_+-r_-:\,\gtgr \rightarrow \gtg$ is an isomorphism of
linear spaces. In that case \gtg\ is called factorizable
and any $X\in \gtg$ has a unique decomposition
\eq 
X=X_+-X_-,
\en
with $(X_-, X_+) \in \mbox{Im}(i_r)$.
The map $i_r$ gives rise to a Lie group embedding 
$I_r:G_r=G_-\times G_+\rightarrow G\times G$. Followed 
by the group inversion in the first factor and a 
subsequent group multiplication of factors it 
defines a local homeomorphism on the Poisson Lie groups $G_r$ and $G$. 
This means that in the neighborhood of the group 
identity any group element $g\in G$ admits a unique decomposition
\eq
g=g_-^{-1}g_+,\label{cfac}
\en
with $(g_-,g_+)\in \mbox{Im}(I_r)$.

The group manifold equipped with Sklyanin bracket 
(\ref{Sklyanin}) plays the role of the phase space 
for a classical dynamical system governed by a 
Hamiltonian constructed in the following way. 
Let $\phi$ by an automorphism of \gtg\ which 
preserves the classical $r$-matrix
\eq
(\phi\otimes \phi)r=r
\en
and which defines an automorphism of $G$ denoted by the same symbol.
The Hamiltonian $h$ is taken as any smooth function on 
$G$ invariant with respect to the twisted conjugation. This means that
\eq
h(g)=h(g_1^{\phi} g (g_1)^{-1}),\,\,\, g^{\phi}\equiv\phi(g)\label{twconj}
\en
holds for any two group elements $g,g_1\in G$.
The functions satisfying (\ref{twconj}) are in involution 
with respect to the Sklyanin bracket, so they play the role
of integrals of motion for the 
dynamical system on $G$ just described above.
We shall denote this involutive 
subset of $C^{\infty}(G)$ as $I^{\phi}(G)$.

For any smooth function $f$ on $G$ let us introduce $D_f(g) \in \gtgr$
by the following equality
\eq
\langle D_f (g),X\rangle=(d/dt)_{t=0}f(ge^{tX}).
\en

We shall also use the symbol $\nabla_f(g)$ 
for the corresponding element of 
\gtg
\eq 
\nabla_f(g)=(D_f(g))_+-(D_f(g))_-
\en
and we shall refer to it as to the gradient of $f$.
If $h_1, h_2\in I^{\phi}(G)$ then the corresponding gradients commute.

Now we can formulate the main classical theorem: 
\begin{thm}[main classical theorem] \label{main}
(i) Functions $h\in I^{\phi}(G)$ are in involution with 
respect to the Sklyanin bracket on $G$. 
(ii) The equations of motion defined by Hamiltonians 
$h\in I^{\phi}(G)$ are of the Lax form
\eq 
dL/dt = \phi(M^{\pm})L -LM^{\pm}, \label{clax}
\en
with $M^{\pm}=(D_h)(L)_{\pm}$ and $L\in G$.
(iii) Let $g^{\pm}(t)$ be the solutions to the 
factorization problem \/\emph{(\ref{cfac})} 
with the left hand side given by
\eq
g(t)=\mbox {exp}(t\nabla_h(L(0))).
\en 
The integral curves of equation \/\emph{(\ref{clax})} are given by
\eq
L(t)=\phi(g_{\pm}(t))L(0)g_{\pm}(t)^{-1}.\label{csl}
\en
\end{thm}

We shall not to prove this theorem. Interested reader can consult
\cite{STSDress}, \cite{ReySTS} for the proof. We just mention that 
there is an easy direct proof
and a more conceptual one, which reflects the 
all rich structure of the theory of Poisson Lie 
groups and the geometry related to the theory of 
integrable systems. The strategy of the second proof 
is to show that the Lax equations 
are obtained by a change of variables from the simplest 
$G_-\times G_+$-invariant Hamiltonian systems on the so-called 
classical Heisenberg double
(the Poisson Lie generalization of the of the 
cotangent bundle $T^*(G)$). 
This construction allows also, using the 
Poisson Lie variant of the symplectic reduction, 
to give one more description of the symplectic leaves 
of $G$ which are known 
to be the orbits of the dressing action of $G_r$ on $G$ \cite{STSDress}.

It is often useful to consider the 
Lax equations (\ref{clax}) corresponding to
different Hamiltonians from 
$I^{\phi}(G)$ simultaneously, using different
time parameters corresponding to the different Hamiltonians.
Usually there is given
a hierarchy (a complete set) of functionally independent
Hamiltonians $h_{\alpha}$, with $\alpha$ 
running over some label set $I$.
The corresponding time parameters are
$\{t_{\alpha}\}_{\alpha \in I}\equiv{\bf t}$
and we index by $\alpha$ also the corresponding gradients and matrices
$M^{\pm}$ entering the Lax equations. 
Then we obtain the following equations
for $L({\bf t})\in G$, $M^{\pm}_{\alpha}({\bf t})\in \gtg$
and $M_{\alpha}\equiv \nabla_{h_{\alpha}}\in \gtg$
\eq
\partial L/\partial t_{\alpha} 
= \phi(M_{\alpha}^{\pm})L -LM_{\alpha}^{\pm}, \label{1a}
\en
\eq
\partial M_{\alpha}/\partial t_{\beta} = [M^{\pm}_{\beta}, M_{\alpha}]  \label{1b}
\en
and
\eq
\partial M^{\pm}_{\alpha}/\partial t_{\beta} 
-\partial M^{\pm}_{\beta}/\partial t_{\alpha}
= [M^{\pm}_{\beta}, M^{\pm}_{\alpha}]. \label{1c}
\en
Here the integral curves $L({\bf t})$ corresponding 
to the commuting dynamical flows on $G$ are given as 
above by the twisted conjugation like in (\ref{csl})
with the factors of
\eq
g({\bf t})=\mbox {exp}(\sum_{\alpha}t_{\alpha}M_{\alpha}(0)). \label{1d}
\en
In the case of a concrete dynamical system, 
we pick up a proper symplectic leaf
on $G$. Usually it is taken in a way 
that $I^{\phi}(G)$ contains 
enough first integrals to ensure that the system is 
completely integrable (in a proper sense).
The group element $g({\bf t})$ can then be brought 
by a similarity transformation to the form
\eq g({\bf t})=\varphi(0)\mbox {exp}(\sum 
 t_{\alpha}X_{\alpha})\varphi(0)^{-1},
\en
where $X_{\alpha}$ are generators of some abelian subalgebra of $G$,
so that the group element $g({\bf t})$ describes an 
embedding of one of the commutative subgroups of $G$ 
into $G$ itself.

Let us now consider the 
quantization of the Lie-Poisson structure. 
The reader is referred to the existing monographs 
on quantum groups (e.g. \cite{ChP}) for the necessary 
information on quantum groups.

\subsection{Heisenberg equations of motion}

As quantized phase space we take the 
non-commutative Hopf algebra $F$
of functions on a quantum group, dual via the pairing
$\la.,.\ra:\, U \ot F \rightarrow k$ to a 
quasitriangular Hopf algebra $U$.
The quantum analog of Sklyanin's bracket is then \cite{FRT}
\eq
R T_1 T_2 = T_2 T_1 R. \label{RTT}
\en
The $R$-matrix can be expanded as 
$R = 1 + h r + {\cal O}(h^2)$, where $h$
is a deformation parameter that 
gives the correspondence to classical
mechanics:
\[
\frac{[f,g]}{h} = \{f,g\} \quad (\mbox{mod } h),
\]
such that for instance
$
0 = R T_1 T_2 - T_2 T_1 R 
= [T_1 , T_2] \quad (\mbox{mod } h)
$
and
\[
0 = \frac{R T_1 T_2 - T_2 T_1 R}{h} 
= \{T_1,T_2\} - [T_1 T_2, r] \quad (\mbox{mod } h) ,
\]
\emph{i.e.} 
$\{T_1,T_2\} = [T_1 T_2, r]$ is the
classical limit of (\ref{RTT}) as desired. 
(Note that (locally) all integrable systems have such a
classical $r$-matrix \cite{BaVi}).

In these and the following expressions $T$ 
may either be interpreted simply
as a Matrix $T \in M_n(F)$ or, much more general, 
as the canonical element of $U \ot F$ : \
Let $\{e_i\}$ and $\{f^i\}$
be dual linear bases of $U$ and $F$ respectively. 
It is very convenient to work with the canonical
element in $U \ot F$ (also called the universal $T$-matrix \cite{FRT}, 
for an elementary overview see
\cite{CSW}),
\eq
T = \sum_i e_i \ot f^i \quad \in \quad U \ot F,
\en
because equations expressed in terms of it will be
reminiscent of the familiar expressions
for matrix representation---but we still keep 
full Hopf algebraic generality.
For the same reason we will often write ``$T_1$'' in
place of ``$T_{12}$'' when we use a notation
that suppresses direct reference to the second tensor
space of $T$; multiplication in $F$ is understood
in that case. Example:
$R_{12} T_1 T_2 = T_2 T_1 R_{12}$
is short for
$R_{12} T_{13} T_{23} = T_{23} T_{13} R_{12}$.

The set of cocommutative elements of $F$ form a
commutative subalgebra $I \subset F$ \cite{DrQG}. 
If we choose a Hamiltonian
from this set, it will commute with all other cocommutative
elements, which will consequently be constants of motion.
This observation can be generalized to twisted cocommutative
elements: Let $\phi$ be an automorphism of $U$ that 
preserves the universal $R$-matrix,
\eq
(\phi \ot \phi)(R) = R. \label{phiR}
\en
($\phi$ is the quantum analog of an 
automorphism of a Poisson-Lie Group.)
Let $\phi^*$ be the pullback of $\phi$ to $F$, \ie\
$\langle x , \phi^*(f) \rangle = \langle \phi(x) , f
\rangle$.
The Hamiltonian $h \in F$ shall be a twisted cocommutative 
function,\footnote{Quantum traces are 
twisted cocommutative functions
with a non-trivial $\phi$ given by the square of the 
antipode, while ordinary traces are simply cocommutative.}
\ie
\eq
\Delta' h = (\phi^*\otimes\id)(\Delta h), \label{phiDelta}
\en
where $\Delta' = \tau \circ \Delta$ is the opposite
comultiplication. The set of twisted 
cocommutative functions also form a
commutative subalgebra $I^\phi \subset F$.

In the following we shall 
present the more general twisted case.
The untwisted formulation can be obtained by omitting ``$\phi$''
in all expressions.
The dynamics of our system is governed by the Heisenberg
equations of motion
\eq
i \dot f = [h , f], \quad \forall f \in F.
\en
These can equivalently be written in terms of the universal
$T$ as
\[i\dot T \equiv i\sum e_i \ot \dot f^i = [\id \ot h , T] 
= \langle T_{13} T_{23} - T_{23} T_{13} , 
h \ot \id^2 \rangle,\] 
or short
\eq
i \dot T_2 =  \langle [T_1 , T_2] , h \ot \id \rangle.
\en
Our strategy to solve this equation will be to embed
the quantized phase-space
$F$ into a bigger algebra (the Heisenberg double, 
$D_H \approx F \ot U$ as a vector space),
where the equations take on a particularly simple form:
The image under this embedding of a 
Hamiltonian $h \in I^\phi$ 
is a casimir in $U$ which leads to trivial time evolution
in $U \subset D_H$ and simple (linear) evolution in 
$D_H$. Projecting the solution 
back to $F$ 
we will find that Heisenberg's equations can be written
in (twisted) Lax form and our original problem
is solved by factorization just
like in the classical case.

\section{Heisenberg double with twist}

The Heisenberg Double of $F$ shall refer to the the semi-direct product
algebra $D_H = F \semi U$ \cite{Sweedler,AF,STS}. 
It is also known as the quantum
algebra of differential operators \cite{Zum,SWZ} or the quantum cotangent
bundle on $F$; see \emph{e.g.} \cite{Fr} for the corresponding classical
Heisenberg double. $D_H$ is isomorphic to $F \ot U$ as a vector
space; it inherits the product structures of $F$ and $U$;
mixed products are obtained from the left action of $U$ on $F$.
All relations can be conveniently summarized
in terms of the canonical element $T$ of $F \ot U$~\cite{CSW}:
\eq
T_{23} T_{12} = T_{12} T_{13} T_{23}. \label{HD} \hfill\mbox{\it
Heisenberg Double}\qquad
\en
This equation gives commutation relations for
elements  $x \in U$ with elements $f \in F$
that equip $F \ot U$ with an algebra structure:
$\: x\cdot f
= \la\id \ot \Delta x,\Delta f\ot \id \ra\:\in\: F \ot U$.
In the setting of interest to us, $F$ is a co-quasitriangular Hopf
algebra who's structure (\ref{RTT}) is determined by a universal
$R$-matrix. Following Drinfel'd's construction \cite{DrQG}
we shall assume that $U$ is itself the quantum double of a Hopf
algebra $U_+$; the universal $R$ arises then
as the canonical element in $U_- \ot U_+$, where
$U_- = U_+^{* \mbox{\tiny op} \Delta}$.
The Yang-Baxter equation
\eq R_{23} R_{13} R_{12} = R_{12} R_{13} R_{23} 
\hfill\mbox{\it
YBE -- Quantum Double}\qquad \label{YBE}\en
plays the same role for
the quantum double $U = U_+ \Join U_-$
as (\ref{HD}) plays for the Heisenberg Double.
The spaces $U_+$ and $U_-$ are images of the
two mappings $R^\pm: F \rightarrow U_\pm$ associated with the universal 
$R \in U_- \ot U_+$:
\eq
R^+(f) = \la R_{21} , \id \ot f\ra, 
\qquad R^-(f) = \la R\i_{12}, \id \ot f\ra.
\en

The twisted Heisenberg double was introduced in \cite{STS}.
The most convenient description for our purposes
of the dual Hopf algebra $U$ with twist is in terms of the universal 
invertible twisted-invariant 2-tensor $Y \in U \ot U$:
\eq
Y_1 R\p_{12} Y_2 R_{21} = R_{12} Y_2 R\p_{21} Y_1. \label{YY}
\en
Here as well as in the following the superscript ${}^\phi$ denotes
the application of the automorphism $\phi$ to the first tensor space:
$R^\phi \equiv (\phi\ot\id)(R)$.
Twisted invariance means:
\eq
Y_{12} T\p_1 T_2 =  T_1 T_2 Y_{12}.
\en 
The mixed relations
\eq
Y_1 T_2 = T_2 R_{21} Y_1 R\p_{12} \label{YT}
\en
complete the description of the Heisenberg double.
(In the case of a trivial twist $\phi = \id$ we may chose
$Y = R_{21} R_{12}$; equation (\ref{YY}) is then a consequence
of the Yang-Baxter equation (\ref{YBE}).)

Crucial for the construction that we are going to present 
is that $T$ factorizes in $U \ot F$ as \cite{FRT} 
\eq
T = \Lambda Z,\quad\mbox{with}\quad \Lambda \in U_- \ot F_-,\quad
Z \in U_+ \ot F_+ ,\quad F_\pm = (U_\pm)^* .
\en
We have $U = U_- \ot U_+$ as a linear space \emph{and coalgebra}
and $F = F_- \ot F_+$ as a linear space \emph{and algebra:}
\eq
(\Delta\ot\id)(\Lambda) = \Lambda_1 \Lambda_2, \quad
(\Delta\ot\id)(Z) = Z_1 Z_2, \quad
\Lambda_1 Z_2 = Z_2 \Lambda_1 .
\en
The universal elements $Z$ and $\Lambda$ of $U_\pm \ot F_\pm$
define projections $F \rightarrow F_\pm$ and
$U \rightarrow U_\pm$ that can be used to extract the $F_\pm$
parts of any element of $F$ and the $U_\pm$ parts of
$(-+)$-ordered expressions
in $U$. Let us denote by $(L^+)\i \in U \ot U_+$ and $L^- \in U \ot U_-$
the corresponding images of $Y\i$, such that $Y = L^+ (L^-)\i$.
Using the maps based on $Z$ and $\Lambda$ we can derive a
host of relations between $Y$, $Z$, $\Lambda$ and $L^\pm$:
\begin{prop} The Heisenberg Double is defined by relations
(\ref{RTT}), (\ref{YY}) and (\ref{YT}).
The following relations are consequences of
these and (\ref{HD}):
\eq
Z_1 Y_1 Z_2 = Z_2 Z_1 Y_1 R^\phi_{12} \label{YZ}
\en
\eq
R_{12} Z_1 Z_2 = Z_2 Z_1 R_{12} \label{ZZ}
\en
\eq
R_{12} \Lambda_1 \Lambda_2 = \Lambda_2 \Lambda_1 R_{12}
\en
\eq
Z_1 L^+_1 Z_2 = Z_2 Z_1 L^+_1
\en
\eq
Z_1 L^+_1 \Lambda_2 = \Lambda_2 R_{21} Z_1 L^+_1
\en
\eq
L^-_1 Z_2 = Z_2 R\p_{12}{}\i L^-_1
\en
\eq
L^-_1  \Lambda_2 = \Lambda_2 L^-_1
\en
\eq
R_{12} L^\pm_2 L^\pm_1 = L^\pm_1 L^\pm_2 R_{12}
\en
\eq
R\p_{12} L^+_2 L^-_1 = L^-_1 L^+_2 R_{12}
\en
\eq
Z_{23} \Lambda_{12} = \Lambda_{12} Z_{23}
\en
\eq
Z_{23} Z_{12} = Z_{12} Z_{13} Z_{23}.
\en
\end{prop}
\emph{Remark: } Similar relations involving $R$, $L^\pm$ and $T$ in the presence
of twisting were proposed in \cite{STS}.
Further relations involving $Y$ and $T$ can be easily obtained from
the ones given above. 
The apparent asymmetry between relations involving $Z$ versus those
involving $\Lambda$ is due to our choice of factorizing $T$ as $\Lambda Z$.
We could have also based our analysis on $T = S V$ with $S \in U_+ \ot F_+$
and $V \in U_- \ot F_-$; this would restore the $+$/$-$-symmetry.\\[1ex]
{\bf Proof: } We shall only proof relation (\ref{YZ}) that we are
going to use extensively in the next sections.
(The other relations follow similarly; see also \cite{JSch2}
and the discussion in \cite{JSch}.)\\
The second tensor space of relation (\ref{YT}) is not $(U_-U_+)$-ordered
so we have to resort to a trick:
with the help of (\ref{RTT}) we can derive a new relation
\[
T_1 Y_1 T{}_{\stackrel{\scriptstyle 2}{\scriptscriptstyle -+}}
= R{}_{\stackrel{\scriptstyle 2}{\scriptscriptstyle -}}{}_1 
T{}_{\stackrel{\scriptstyle 2}{\scriptscriptstyle -+}} 
T_1 Y_1 R\p{}_1{}_{\stackrel{\scriptstyle 2}{\scriptscriptstyle +}}
\] 
whose second tensor
space is $(U_-U_+)$-ordered as is easily verified. Projecting to
$U_+$ we obtain $\: T_1 Y_1 Z_2 = Z_2 T_1 Y_1 R\p_{12}\:$ which
can be simplified using $T_1 = \Lambda_1 Z_1$ and $\Lambda_1 Z_2 = Z_2
\Lambda_1$ to yield (\ref{YZ}). \hfill $\Box$

\subsubsection*{A remark on quantum traces and twisting}

We have argued that the cocommutative 
elements of $F$ are natural candidates
for Hamiltonians. Classically 
cocommutativity is equivalent to ad-invariance,
so it would also be natural to 
look for Hamiltonian functions in the
quantum case that are invariant under the
quantum adjoint coaction\footnote{This 
holds for instance for 
quantum traces.}
\eq
\Delta^{Ad}(h) \equiv h_{(2)} \ot S(h_{(1)}) h_{(3)}
= h \ot 1,\label{adinv}
\en
It turns out that both these and the cocommutative 
Hamiltonians are
treated on equal footing 
in the twisted formulation: Requirement (\ref{adinv}) is 
equivalent to
\eq
\Delta h = (\id \ot S^2)(\Delta' h),
\en
\ie\ corresponds to a twisted cocommutative function with the
pullback of the twist
$\phi^*$ given by the square of the antipode.
The twist $\phi = S^2$ is here generated via conjugation by 
an element $u \in U$:
\eq
S^2(x) = u x u^{-1},\quad\forall x \in U.
\en
It seems interesting to study the general case of a twist $\phi$ that
is given via conjugation by some element $\varphi$, \ie
\eq
\phi(x) = \varphi x \varphi\i, \qquad (\varphi\ot\varphi) R = R 
(\varphi\ot\varphi).\label{varphi}
\en
If $\varphi \in U$ then
$\phi^*(f) = \la\varphi 
, f_{(1)}\ra f_{(2)} \la\varphi\i , f_{(3)}\ra$
for all $f \in F$.
(Here we see by the way that $\phi^*(h) = h$ 
holds both for cocommutative
and twisted cocommutative $h$.)
Due to (\ref{varphi}), 
$f \mapsto \la \id \ot \varphi\i , \Delta f\ra$ defines an
algebra isomorphism of $F$, that maps
$I \subset F$ to $I\p \subset F$, \ie\
cocommutative elements to twisted cocommutative
elements.

It is easily verified that all expressions 
containing $\phi$, \emph{e.g.}
(\ref{YY}), (\ref{YZ}), \emph{etc.} continue to hold 
if we omit $\phi$ \emph{and}
replace $Y$ by $Y\cdot(\varphi\ot 1)$. 
Examples:
\eq
R_{21} Y_1 \varphi_1 R_{12} Y_2 \varphi_2 
= Y_2 \varphi_2 R_{21}
Y_1 \varphi_1 R_{12},
\en
\eq
Z_1 Y_1 \varphi_1 Z_2 = Z_2 Z_1 Y_1 \varphi_1 R_{12}, \qquad etc.
\en
This gives a nice mnemonic for where
to put the $\phi$'s---\emph{even when an element $\varphi$ does
not exist in $U$}: First we write 
expressions without  $\phi$, then we
formally 
replace all $(L^-)^{-1}$'s by 
$(L^-)^{-1}\cdot(\varphi\ot 1)$ (and consequently 
$Y$ by $Y\cdot(\varphi\ot 1)$),
finally we
remove all $\varphi$'s from the 
expression with the help of
relation (\ref{varphi}).
\ \emph{Remark:}
$Y = L^+ (L^-)^{-1}$ but $Y \neq R_{21} R_{12}$ in the
twisted case. In case we know an element $\varphi$ that
satisfies (\ref{varphi}), we can realize $L^\pm$
in terms of the universal $R$ for instance as $L^+ = R_{21}$
and $L^- = \varphi_1 R\i_{12}$. 
There is however some remaining ambiguity in this choice.

\subsection{Embedding the operator algebra into the double}
\label{s:embed}

Here we will show how to embed $F$ into $D_H$ 
in such a way that any (twisted)
cocommutative element of $F$ is mapped to a 
casimir operator of $U \subset
D_H$.
\begin{prop}
The following element of $U \ot D_H$ 
\eq
\widetilde T = \phi(Z) Y^{-1} Z^{-1},
\en
where $\phi(Z) \equiv (\phi\ot\id)(Z)$, satisfies 
\eq
R_{12} \widetilde T_1 \widetilde T_2 
= \widetilde T_1 \widetilde T_2 R_{12}
\label{RTTtilde}
\en
and thus defines an embedding of $F \hookrightarrow D_H:\:
f \mapsto \langle \widetilde T , f \ot \id \rangle$,
that is an algebra homomorphism. (The picture
of $F$ in $D_H$ by this embedding will be denoted $\widetilde F$.)
\end{prop}

\paragraph{Proof:} Start with (\ref{YY}) in form
$Y\i_1 R^\phi_{21}{}\i Y_2\i 
= R_{21}\i Y_2\i R\p_{12}{}\i Y_1\i R_{12},$
multiply by $Z\i_2 Z\i_1$ from the right and use (\ref{ZZ}) to obtain
\[
Y\i_1 Z\i_1 Y\i_2 Z\i_2 = R_{21}\i Y\i_2 \underline{R\p_{12}{}\i Y\i_1
Z\i_1 Z\i_2} R_{12}.
\]
Applying equation (\ref{YZ}) to the underlined part gives
\eq
Y\i_1 Z\i_1 Y\i_2 Z\i_2 = R_{21}\i Y\i_2 Z\i_2 Y\i_1 Z\i_1 R_{12}
\label{zwischen}
\en
and as a corollary:
$R_{12} Z_2 Y_2 Z_1 Y_1 = Z_1 Y_1 Z_2 Y_2 R_{21}$. 
Now use equation (\ref{YZ}) twice: once in the form
$
Z_1 Y_1 \phi(Z_2) = \phi(Z_2) Z_1 Y_1 R_{12},
$
which follows from $(\phi\ot\phi)(R) =R$,
to replace $Y\i_1 Z\i_1$ on the LHS of (\ref{zwischen})
and once to replace $R_{21}{}\i Y\i_2 Z\i_2$ on the RHS of (\ref{zwischen}).
Multiplying the resulting expression
by $\phi(Z_i)$ from the left and using (\ref{ZZ}) in the form
$
R_{12} \phi(Z_1) \phi(Z_2) = \phi(Z_1) \phi(Z_2) R_{12}
$
gives our result (\ref{RTTtilde}). \hfill$\Box$

\begin{prop} 
The image $\widetilde h$ of the Hamiltonian $h$
under the embedding $F \rightarrow \widetilde F$ is a casimir in
$U \subset D_H$. We can find the following explicit expression:
\eq
\widetilde h = \la \widetilde T , h \ot \id \ra
  = \la u_1\i Y_1\i , h \ot \id \ra, \label{casimir}
\en
where\footnote{Here and in the following we
will use the following convenient notation that was 
brought to our attention
by C.\ Chryssomalakos: The second subscripts denote 
the \emph{order of
multiplication} in a given tensor space. 
Consider for example
$R = \sum_i \alpha_i \ot \beta^i$, then
$(S^2 \ot \id)(R)_{1_21_1}$ equals $\sum_i \beta^i S^2(\alpha_i)$ and lives in
tensor space 1.}
$u\i = (S^2 \ot \id)(R)_{1_21_1}$ and
satisfies $u\i x = S^2(x) u\i, \ \forall x \in U$ .
\end{prop}

\paragraph{Proof:} We have to proof two things: 
1) $\widetilde h$ commutes with
all elements of $U$ and 2) $\widetilde h$ is an element of $U$ with the 
given expression.\\[1ex]
Ad 1): Here is a nice direct calculation that shows that $\widetilde h$ 
commutes with $Y^{-1}$
and hence (in the factorizable case) with all of $U$:\\[1ex]
Start with the twisted reflection equation (\ref{YY}) in the form
\[
Y\i_1 \underline{R\p_{21}{}\i Y\i_2} 
= R_{21}\i Y\i_2 R\p_{12}{}\i Y\i_1 R_{12},
\]
apply (\ref{YZ}) with subscripts 1 and 2 exchanged 
to the underlined part,
rearrange and multiply by $\phi(Z_1)$ from the left to obtain:
\[
\phi(Z_1) Y\i_1 Z\i_1 Y\i_2 
=\phi(Z_1) R_{21}\i Y\i_2 R\p_{12}{}\i Y\i_1 R_{12} Z\i_2 Z\i_1 Z_2.
\]
Now we can use (\ref{YZ}) twice, first in the form
$\phi(Z_1) R_{21}\i Y\i_2 = Y\i_2 Z\i_2 \phi(Z_1) Z_2$ and then
in the form $Z_2 R\p_{12}{}\i Y\i_1 = Y\i_1 Z\i_1 Z_2 Z_1$, to 
remove two $R$'s from the RHS. The resulting expression,
simplified with the help of (\ref{ZZ}), is
\[
\phi(Z_1) Y\i_1 Z\i_1 Y\i_2 
= Y\i_2 Z\i_2 \phi(Z_1) Y\i_1 Z\i_1 R_{12} Z_2.
\]
Contracting with $h$ in the first tensor space and using
$h_{(1)} \ot\ldots\ot h_{(4)} 
= h_{(2)}\ot\ldots\ot h_{(4)}\ot\phi^* h_{(1)}$,
which follows from the twisted 
cocommutativity of $h$, we can move $R_{12}$
three places to the left:
\[
\la\phi(Z_1) Y\i_1 Z\i_1 Y\i_2, h \ot \id \ra
= \la Y\i_2 Z\i_2 R\p_{12} \phi(Z_1) 
\underline{Y\i_1 Z\i_1 Z_2},h \ot \id\ra.
\]
Applying (\ref{YZ}) once more to 
the underlined part and simplifying the
resulting expression
with the help of (\ref{ZZ}) in
the form $R\p_{12} \phi(Z_1) Z_2 
= Z_2 \phi(Z_1) R\p_{12}$ we finally obtain
$
\la\phi(Z_1) Y\i_1 Z\i_1 Y\i_2, h \ot \id \ra
=\la Y\i_2 \phi(Z_1) Y\i_1 Z\i_1, h \ot \id \ra,
$
\ie\
\eq
[1 \ot \widetilde h, Y\i ] = 0 .
\en
\vspace{1ex}
Ad 2): 
Now we will derive the explicit expression for $\widetilde h$. 
(Using that expression
it is also possible to prove that $\widetilde h$ is a casimir
in $U$.)
We start with equation (\ref{YZ}), written as
$
Z_2 R\p_{12}{}\i Y\i_1 Z\i_1 = Y\i_1 Z\i_1 Z_2,
$
and move the $R$ to the RHS with the help of
its opposite inverse 
$\bar R\p \equiv (S^2 \ot \id)(R\p)$, which satisfies 
$\bar R\p_{12_2} R\p_{12_1}{}\i = 1 \ot 1$. We find
$
Z_2 Y\i_1 Z\i_1 = \bar R\p_{12_2} Y\i_1 Z\i_1 Z_{2_1}.
$
Let us now multiply tensor spaces 1 and 2 so that the
two $Z$'s on the RHS cancel
\[
Z_{1_3} Y\i_{1_1} Z\i_{1_2} 
= \bar R\p_{1_11_5} Y\i_{1_2} Z\i_{1_3} Z_{1_4}
= \bar R\p_{1_11_3} Y\i_{1_2}.
\]
If we now contract this expression with $h$ in the first tensor space,
we can use the twisted cocommutativity of $h$ in the form
$h_{(1)} \ot h_{(2)} \ot h_{(3)} 
= h_{(2)} \ot h_{(3)} \ot \phi^* h_{(1)}$
to change the order of multiplication in the first tensor space
on both sides of the equation:
\[
\la \phi(Z_{1_1}) Y\i_{1_2} Z\i_{1_3},h \ot \id\ra
= \la \bar R_{1_21_1} Y\i_{1_3},h \ot \id\ra,
\]
\ie\ 
$\la \widetilde T , h \ot \id \ra
  = \la u_1\i Y_1\i , h \ot \id \ra$.
This is precisely the expression (\ref{casimir}) 
that we wanted to prove.

We would like to briefly sketch how to prove that $\widetilde h$ is a
casimir starting from (\ref{casimir}):
$h' = \la u\i \ot \id , \Delta(h)\ra$ is an element of $F$ which is
coinvariant with respect to the twisted adjoint action. 
 (This follows from twisted cocommutativity of $h$
and the fact that $u\i$ generates $S^2$).
$Y$ on the other hand is a twisted invariant 2-tensor in $U \ot U$.
Being the contraction of $Y$ by $h'$, 
$\widetilde h$ is itself an ad-invariant
element of $U$ and hence a casimir operator: 
$T_2 \la Y\i_{12} , h' \ot \id\ra
= \la T\p_1{}\i Y\i_{12} T_1 , h' \ot \id\ra T_2
= \la Y\i_{12} , h' \ot \id\ra T_2$ .
\hfill$\Box$

\subsection{Dynamics in the double}

Now that we have found the image of the Hamiltonian under the
embedding of the quantized phase space $F$ into the Heisenberg double
we can study Heisenberg's equations of motion in the double.
These are
\eq
i \dot\mathcal{O} 
= [\widetilde h , \mathcal{O}], \quad \forall \mathcal{O} \in D_H.
\label{heom}
\en
Time evolution in the $U$-part of $D_H$ 
is trivial (because $\widetilde h$ 
is central
in $U$)
\eq
i \dot x = [\widetilde h , x] = 0, 
\quad \forall x \in U \subset D_H.\label{uconst}
\en
In the $F$-part we find simple linear equations
\eq
i \dot T = [1\ot\widetilde h , T] 
= T (\Delta \widetilde h - 1 \ot \widetilde h) =: T \xi
\label{Tdot}
\en
that are solved by exponentiation because $\xi$ 
is an element of $U\ot U \subset U \ot D_H$ and hence
time-independent, see
(\ref{uconst}),
\eq
T(t) = T(0) e^{-i t \xi}.
\en
Here are some alternative useful expressions for 
$\xi = \Delta \widetilde h - 1 \ot \widetilde h$:
Equation (\ref{Tdot}) slightly rewritten gives
\eq
T_2 \xi_2 = \la\widetilde T_1 T_2 - T_2 \widetilde T_1 , h \ot \id\ra.
\label{txi}
\en
Starting from (\ref{YT})
one can derive
\eq
\xi = \left\la u_1\i(R\p_{12}{}\i Y\i_1 R_{21}\i - Y\i_1) 
              , h \ot id\right\ra
\en
and
\eq
\xi = \left\la \left((S\i \circ \phi \ot \id)(R_{12} R_{21})
      - 1 \right) u\i_1 Y\i_1 , h \ot \id\right\ra.
\en	  
We have thus far been able to give the explicit time evolution in the
Heisenberg double. In section~\ref{s:qlax} we will come closer
to the solution to
the original problem---Heisenberg's equations of motion---via
explicit expressions for the evolution of $\widetilde T(t)$.

\section{Quantum Lax equation}
\label{s:qlax}

We will now derive an explicit expression for the time evolution 
of $\widetilde T$.
Using the time-independence of 
$Y\i \in U \ot U \subset U \ot D_H$ we find
\eq
\widetilde T(t) = \phi(Z(t)) \, Y\i \, Z\i(t) = 
\phi(Z(t) Z\i(0)) \, \widetilde T(0) \, Z(0) Z\i(t). \label{Ttfirst}
\en
If we had started with an alternative $\widetilde T$ expressed
in terms of $\Lambda$ and $Y$ 
we would have found an expression involving $\Lambda$ instead of $Z$.
Such considerations lead to the following proposition:
\begin{prop}
Let \quad$\widetilde g_+(t) = Z(t) Z(0)\i$\,,\quad
$\widetilde g_-(t) = \Lambda\i(t) \Lambda(0)$\quad and
$\widetilde M^\pm(t) = i \,\dot{\widetilde g}_\pm(t) 
\widetilde g_\pm\i(t)$.
The time-evolution of $\widetilde T$ is given via conjugation
by
\eq
\widetilde g_\pm(t) = \exp(-it(1 \ot \widetilde h)) 
\exp(it(1 \ot \widetilde h - \widetilde M_\pm(0))) : \label{gpt}
\en
\eq
\widetilde T(t) = \phi(\widetilde g_+(t)) \widetilde T(0)
                  \widetilde g_+(t)\i 
                = \phi(\widetilde g_-(t)) \widetilde T(0)
                  \widetilde g_-(t)\i 
\label{Tt}
\en
and Heisenberg's equation of motion can be written in Lax form
\eq
i \,\frac{d}{dt}\widetilde T = \phi(\widetilde M^+) \widetilde T 
- \widetilde T \widetilde M^+
= \phi(\widetilde M^-) \widetilde T 
- \widetilde T \widetilde M^- . \label{mttm}
\en
\end{prop}
\paragraph{Proof:} The definition of $\widetilde M^\pm(t)$
can be used to express $\widetilde g_\pm(t)$
in terms of $\widetilde M^\pm(t)$. From (\ref{heom}) we have
\[
i \,\frac{d}{dt}\widetilde g_\pm(t) 
=\widetilde M_\pm(t) \widetilde g_\pm(t) 
= e^{-it(1 \ot \widetilde h)} \widetilde M_\pm(0) e^{i t(1 \ot
\widetilde h)} \widetilde g_\pm(t) ;
\]
this can be integrated
with the initial condition $\widetilde g_\pm(0) = 1$ to give
equation~(\ref{gpt}). 
If we differentiate (\ref{Tt}), we find equation (\ref{mttm}).
What is left to proof is equation (\ref{Tt}).
$
\widetilde T(t) = \phi(\widetilde g_+(t)) \widetilde T(0)
\widetilde g_+(t)\i 
$
is simply (\ref{Ttfirst})
expressed in terms of $\widetilde g_+(t)$.
The time evolution in $D_H$ is an algebra 
homomorphism and so we can decompose
$T(t)$ = $\Lambda(t) Z(t)$ with $\Lambda(t) \in U_- \ot F_-(t)$ and 
$Z(t) \in U_+ \ot F_+(t)$.
It is now easy to see that 
$\phi(\widetilde g_+(t)) \widetilde T(0) \widetilde g_+(t)\i 
= \phi(\widetilde g_-(t)) \widetilde T(0) \widetilde g_-(t)\i$     
is equivalent to          
$\phi(T(t)) Y\i T\i(t) = \phi(T(0)) Y\i T\i(0)$, \ie\ we need to show
that $\phi(T) Y\i T\i$ is time-independent: From (\ref{Tdot})
we get
\[
i \, \frac{d}{dt} \left(\phi(T) Y\i T\i\right)
= \phi(T) \left(\phi(\xi)  Y\i - Y\i \xi\right) T\i =0.
\] 
(That this is zero
can be seen from the explicit expression
$\xi = \Delta\widetilde h - 1 \ot \widetilde h$: \quad
$(\phi\ot\id)(\Delta\widetilde h) Y\i - Y\i \Delta\widetilde h = 0$ 
because $Y\i$ is a twisted invariant 2-tensor and 
$[1 \ot \widetilde h , Y\i] = 0$ because 
$\widetilde h$ is a casimir operator.)
\hfill$\Box$\\[1ex]
We will now proceed to derive explicit expressions for 
$\widetilde M^\pm$ in terms of $h$. 
We will not use the expressions for $\xi$
but rather work directly with $Z$ and $\Lambda$. 
First we prove the following lemma:
\begin{lemma} 
The following two relations hold in $U \ot U \ot D_H$:
\eq
\Lambda\i_2 \widetilde T_1 \Lambda_2 
= \widetilde T_1 R_{21}\i, \label{first}
\en
\eq
Z_2 \widetilde T_1 Z_2\i = R\p_{12} \widetilde T_1. \label{second}
\en
\end{lemma}
\paragraph{Proof:}
We need to use 
\eq
Z_1 Y_1 \Lambda_2 = \Lambda_2 R_{21} Z_1 Y_1 ,
\en
\begin{flushleft}
   which follows with $T = \Lambda Z$ from (\ref{YT}) and (\ref{YZ}).
   We have $\Lambda\i_2 \widetilde T_1 \Lambda_2 =
   \Lambda\i_2 \phi(Z_1) 
   \underline{Y_1\i Z\i_1 \Lambda_2} 
    = \Lambda\i_2 \phi(Z_1) \Lambda_2 Y\i_1 Z\i_1
   R_{21}\i = \widetilde T_1 R_{21}\i$ 
   which proves (\ref{first}).
   Similarly:
   $Z_2 \widetilde T_1 Z_2\i = Z_2 \phi(Z_1) 
   \underline{Y\i_1 Z\i_1 Z\i_2} =
   \underline{Z_2 \phi(Z_1) R\p_{12}} Z_2\i Y_1\i Z_1\i 
    = R\p_{12} \widetilde T_1$,
   which proves (\ref{second}). \hfill$\Box$
\end{flushleft}
\begin{prop} 
It holds that
\begin{eqnarray}
\widetilde M^+(t) & = & 1 \ot \widetilde h 
- \la\widetilde T_1(t) R_{12} , h \ot \id\ra\quad\in\quad U_+ \ot
\widetilde F , \\
\widetilde M^-(t) & = & 1 \ot \widetilde h 
- \la \widetilde T_1(t) R\i_{21}  , h \ot \id\ra \quad\in\quad U_- \ot
\widetilde F.\label{mminus}
\end{eqnarray}
\end{prop}
\begin{flushleft}{\bf Proof: }
   $\widetilde M^+(t) = i \dot Z(t) Z(t)\i =
   \displaystyle{
      {d \over d t}\left(e^{-i t (1 \ot \widetilde h)}
      Z(0) e^{i t (1 \ot \widetilde h)}\right) Z(t)\i
   }$
   $=\displaystyle{
        e^{-i t (1 \ot \widetilde h)}
        \left(1 \ot\widetilde h 
        - Z(0)(1 \ot \widetilde h)Z(0)\i\right)
        e^{i t (1 \ot \widetilde h)}
   }$
   $=\displaystyle{1 \ot \widetilde h - 
   e^{-i t (1 \ot \widetilde h)} 
   \la (Z_2 \widetilde T_1 Z_2\i)(0),h \ot \id\ra
   e^{i t (1 \ot \widetilde h)}
   }$
   $=1 \ot \widetilde h - \la R\p_{12} \widetilde T_1(t) , h \ot \id\ra$
   $=1 \ot \widetilde h 
   - \la\widetilde T_1(t) R_{12} , h \ot \id\ra$,\\
   where we have used (\ref{second}) and $(\phi^*\ot\id)(\Delta h) =
   \Delta'h$.\\ The proof
   of (\ref{mminus}) is based on (\ref{first})  
   but otherwise completely analogous. 
   \hfill$\Box$
\end{flushleft}
Just like $T$ factorizes as $T(t) = \Lambda(t) Z(t)$, we shall think of
$\widetilde g_\pm \in U_\pm \ot D_H$ as factors
of an element of $U \ot D_H$:
\eq
\widetilde g(t) = \widetilde g_-\i(t) \widetilde g_+(t) .
\en
\begin{prop}
$\widetilde g(t) \equiv 
\widetilde g_-\i(t) \widetilde g_+(t)$ and its factors $g_-(t)$
and $g_+(t)$ are in fact elements of 
$U \ot \widetilde F \subset U \ot D_H$ as
is apparent from the following expression:
\eq
\widetilde g(t) = Z(0) \exp(-i t \xi) Z\i(0) = \exp(-i t \widetilde M(0)),
\en
where
$
\widetilde M \equiv \widetilde M_+ - \widetilde M_-
= \la \widetilde T_1 (R\i_{21} - R_{12}), h \ot \id\ra
\in U \ot \widetilde F.
$
\end{prop}
\begin{flushleft}{\bf Proof: } $\widetilde g(t) = 
\widetilde g\i_-(t) \widetilde g_+(t) = \Lambda\i(0) T(t) Z\i(0)
= Z(0) \exp(-i t \xi) Z\i(0)$.
{}From equation (\ref{txi}) and 
$T = \Lambda Z$:
$Z \xi Z\i = \Lambda\i_2 \la \widetilde T_1 T_2 
- T_2 \widetilde T_1 , h \ot \id \ra
Z_2\i = \la \Lambda\i_2 \widetilde T_1 \Lambda_2 
- Z_2 \widetilde T_1 Z_2\i , h \ot \id\ra
= \la \widetilde T_1 R\i_{21} - R\p_{12} \widetilde T_1 , h \ot \id\ra = 
\widetilde M_+ - \widetilde M_-$ and hence 
$Z \exp(-i t \xi) Z\i
= \exp(-i t (\widetilde M_+ - \widetilde M_-)) 
= \exp(-i t \widetilde M)$.
\hfill$\Box$
\end{flushleft}
So far we have learned a great deal about the 
equations of motion in the Heisenberg double and their solution. 
We are now ready to go back to our original
problem, \ie\ the formulation of the equations of motion in the quantized
phase space $F$ in terms of quantum Lax equations and their solution by
factorization, thus generalizing what has become known as the ``Main Theorem''
to the realm of quantum mechanics. Let us mention that the Lax equations 
presented in this section formalize and generalize the concrete examples 
known for particular integrable models \cite{IK}, \cite{Skly}, \cite{KS}, 
\cite{Maillet}, \cite{Zhang}, \cite{SoWa}.

\section{Solution by factorization}

Using the fact that the embedding via $\widetilde T$ of $F$
into $D_H$ is
an algebra homomorphism we can drop all the $\widetilde{~~}$'s in the
previous section, thereby projecting the solution of the time evolution of
$\widetilde F$ back to $F$. The result can be summarized in a quantum
mechanical analog of theorem~\ref{main}:

\begin{thm}[main quantum theorem] \label{mainquantum}
\mbox{ }
\begin{enumerate}
\item[(i)] The set of twisted cocommutative functions
$I^{\phi}$ is a commutative subalgebra of $F$. 
\item[(ii)] The equations of motion defined by Hamiltonians
$h\in I^{\phi}$ are of the Lax form
\eq 
i\frac{dT}{dt} = \phi(M^{\pm}) T -T M^{\pm}, \label{qlax}
\en
with 
$\;M^\pm  = \la T_1 (1 - R^\pm)_{21},h \ot\id \ra \in U_\pm \ot F\;$,
 $R^+ \equiv R_{21}$, $R^- \equiv R\i_{12}$
and $T \in U \ot F$.
\item[(iii)] Let $g_{\pm}(t) \in U_\pm \ot F$ be the solutions to the 
factorization problem
\eq
g_-\i(t) g_+(t) =  \exp(-i t M(0))\;\in\; U \ot F , \label{facprob}
\en 
where $M(0) = M^+ - M^-$, then
\eq
T(t)=\phi(g_{\pm}(t)) T(0) g_{\pm}(t)^{-1} \label{qsoln}
\en
solves the Lax equation {\rm (\ref{qlax})}; $\:g_\pm(t)$ are given by 
\eq
g_\pm(t) = \exp(-it(1\ot h))\,\exp(it(1\ot h - M^\pm(0))
\en
and are the solutions to the differential equation
\eq
i  \frac{d}{dt}g_\pm(t) =
M^\pm(t) g_\pm(t), \qquad  g_\pm(0) = 1 . \label{gpm}
\en
\end{enumerate}
\end{thm}
This theorem follows from the geometric construction given in the previous
sections, but we shall also present a direct proof:
\begin{flushleft}{\bf Proof:}\\
(i) Let $f,g \in I\p \subset F$, then $f g \in I\p$. 
Using (\ref{RTT}), (\ref{phiDelta}) 
and (\ref{phiR})
we can show that $f$ and $g$ commute:
\begin{quote}
$f g = \la T_1 T_2, f \ot g\ra
= \la R\i_{12} T_2 T_1 R_{12}, f \ot g\ra
= \la R\i_{13} T_3 T_1 R_{24}, \Delta f \ot \Delta g\ra
= \la R\i_{13} T_3 T_1 \:(\phi \ot \phi)(R_{24}), 
\Delta' f \ot \Delta' g\ra
= \la R\i_{13} T_3 T_1 R_{24}, 
\Delta' f \ot \Delta' g\ra
= \la R_{12} R\i_{12} T_2 T_1, f \ot g\ra = g f$ .
\end{quote}
(ii) From $R^\pm_{21} T_1 T_2 = T_2 T_1 R^\pm_{21}$, (\ref{phiDelta}) 
and (\ref{phiR})  it follows that $\la
T_1\:(\id\ot\phi)(R^\pm_{21}) T_2, h\ot\id\ra 
= \la T_2 T_1 R^\pm_{21} , h\ot\id\ra$
and as a consequence  all terms that contain
$R^\pm_{21}$ cancel on the RHS of 
equation (\ref{qlax}); we are left with
$\phi(M^\pm) T - T M^\pm = \la T_1 T_2 - T_2 T_1 , h \ot \id\ra 
= [ h , T ] = i dT/dt$.\\[1ex]
(iii) The Lax equation (\ref{qlax}) 
follows immediately from (\ref{qsoln}) and
(\ref{gpm}).\\ 
The rest can be proven in three steps:\\ 
a) Let $m_\pm = 1 \ot h - M_\pm$; \quad
$g_\pm(t) = e^{-i t (1\ot h)} e^{i t m_\pm}$ 
are elements of $U_\pm\ot F$ and 
are the solutions to (\ref{gpm}) as can be checked by differentiation: 
$i \frac{d}{dt} g_\pm(t) = 
e^{-i t (1\ot h)}(1\ot h -  m_\pm) e^{i t m_\pm}
= M_\pm(t) e^{i t (1 \ot h)} e^{i t m_\pm} = M_\pm(t) g_\pm(t)$, 
and $g_\pm(0) = 1$\\[1mm]
b) $m_+ = \la T_1 R_{12},h\ot\id\ra$ and 
$m_- = \la T_1 R\i_{21} , h \ot id\ra$ commute:\\
Using (\ref{RTT}), (\ref{YBE}), and
(\ref{phiDelta}) we find
\begin{quote}
$m_+ m_- = \la T_1 R_{13} T_2 R\i_{32}, h \ot h \ot \id\ra
= \la R\i_{12} T_2 T_1 R_{12} R_{13} R\i_{32} , h \ot h \ot \id\ra
= \la R\i_{12} T_2 R\i_{32} T_1 R_{13} R_{12} , h \ot h \ot \id\ra
= \la T_2 R\i_{32} T_1 R_{13} R_{12} R\i_{12} , h \ot h \ot \id\ra
= m_- m_+$
\end{quote}  
{}From a) and b) follows\\[1mm] 
c) $g_-\i(t) g_+(t) =  e^{-i t m_-} e^{i t m_+} = e^{i t (m_+ - m_-)}
= e^{- i t M(0)}$, \ie\ the $g_\pm(t)$ of (\ref{gpm}) solve the
factorization problem (\ref{facprob}). \hfill $\Box$
\end{flushleft}
{\em Remark: } If we replace $R^\pm$ in the definition of $M^\pm$ in the
previous theorem by $(L^\mp)^{-1}$, then the $\phi$'s in (\ref{qlax}) 
and (\ref{qsoln}) will not appear explicitly anymore.

\subsubsection*{Dressing transformations}

We have found two (identical) solutions for the time-evolution
in $F$:
\[
f(t) = \la T(t), f(0) \ot \id\ra , \quad f(0) \in F
\]
with $T(t)$ given in (\ref{qsoln}).
Let us verify that
\[
\phi(g_+(t)) T(0) g_+(t)\i = \phi(g_-(t)) T(0) g_-(t)\i .
\]
Let  $g(t) = g_-(t)\i g_+(t) = \exp(-it M(0))$;  we have to show
that
\eq
\phi(g(t)) = T(0) g(t) T(0)\i  \label{phigt}
\en
which is implied by:
\begin{flushleft}
\begin{quote}
$T M T\i = \la T_2 T_1 (R_{21}\i - R_{12}) T_2\i , h \ot \id\ra
= \la (R_{21}\i - R_{12}) T_1 , h \ot \id\ra
= \la T_1 (\id \ot \phi)(R_{21}\i - R_{12}) , h \ot \id\ra =  \phi(M) .$
\end{quote}
\end{flushleft}
With (\ref{phigt}) we can re-express (\ref{qsoln}) as
\eq
T(t) = \Big( T(0) g(t) T(0)\i \Big)_\pm T(0) \:\Big(g(t)\Big)_\pm\i 
\en
and thus find that the time-evolution in $F$ has the form of a dressing
transformation.
More precisely 
we can identify elements of $F$ 
with elements of $U \ot F$ (via the factorization map):
\eq
e^{i t h} 
\:\mapsto \: e^{i t m_\pm} = R^\pm \cdot e^{i t (1 \ot h)}\cdot (R^\pm)\i 
\:\mapsto \: g = e^{-i t m_-} e^{i t m_+} 
\en
and hence have a map
\eq
F \ni e^{ith}: \: U \ot F \rightarrow U \ot F : \;\; T(0) \mapsto T(t).
\en
Let us choose the same $\pm$-conventions for
$T = \Lambda Z$ and $Y\i = L^- (L^+)\i$ 
as we did for $g = g_-\i g_+$, \ie\ 
$(T)_- = \Lambda\i$, $(T)_+ = Z$,  \emph{etc}.
 We can then write the
embedding of section~\ref{s:embed} in a way that parallels the classical theory:
\begin{eqnarray}
\lefteqn{\phi(T L^+)_- \cdot  (\phi(T) L^-) \cdot (T L^+)_+{}\i }\nonumber\\ 
&& = \phi\left((\Lambda\i)\i Z L^+\right)_- 
      \cdot  (\phi(T) L^-) \cdot  \left((\Lambda\i)\i Z L^+\right)_+ \nonumber\\
&& = \phi(\Lambda\i) \cdot  (\phi(T) L^-) \cdot   (L^+)\i Z\i\nonumber\\
&& = \phi(Z) Y\i Z\i .
\end{eqnarray}
(``$\pm$'' refers to the first tensor space.)\\
\emph{Remark 1: } Note that the multiplication
``$\cdot$'' is
taken in $U \ot D_H$ rather than $U \ot (F \ot U)$; this and the 
form in which the dressing transformations appear in this article are
somewhat non-standard.\\  
\emph{Remark 2: } The formal factorization problem in the case of $U$ 
being factorizable \cite{ReshSTS} remains the same as in the untwisted case.
See Appendix 2 in reference \cite{JSch2}.\\
\emph{Remark 3: }
Also in the quantum case we can consider Lax equations corresponding 
to different twisted cocommutative Hamiltonians simultaneously. 
The equations 
(\ref{1a}), (\ref{1b}), (\ref{1c}) and (\ref{1d}) are still valid. 
\section*{Acknowledgments}
The foundation of this article is previous work done in
collaboration with Michael Schlieker; we thank him for sharing his
ideas on the subject. We are also very grateful to Professor Winternitz
for his interest in this work.


\begin{thebibliography}{article}

\bibitem{Ad} Adler, M.: On a trace formula for pseudo-differential operators
and the symplectic structure of the KdV-type equations. Inv. Math. {\bf 50},
219 (1979)

\bibitem{AF} Alekseev, A. Yu, Faddeev, L. D.: $(T^*G)_t$: 
A Toy model for conformal field theory. Commun. Math. Phys. {\bf 141}, 413 (1991)

\bibitem{BaVi} Babelon, O., Viallet, C. M.: Hamiltonian structures and Lax equations.
Phys. Lett. {\bf B 237}, 411 (1990)

\bibitem{ChP} Chari, V., Pressley, A.: A guide to quantum groups.
Cambridge: Cambridge University Press 1994

\bibitem{CSW} Chryssomalakos, C., Schupp, P., Watts, P.: The Role of the
canonical element in the quantized algebra of differential operators.
Preprint 1993, hep-th/9310100

\bibitem{DrQG} Drinfeld, V. G.: Quantum groups. In Proc. ICM Berkley 1986,
AMS 1987, p. 798

\bibitem{DKN} Dubrovin, B. A., Krichever, I. M., Novikov, P. S.: Integrable
systems. I. In Encyclopedia
of mathematical sciences, Dynamical systems IV,  Berlin, Heidelberg, New York:
Springer 1990

\bibitem{F} Faddeev, L. D.: Integrable models in (1+1)-dimensional field
theory.
In Recent advances in field theory and statistical mechanics, Les Houches,
Section XXXIX, 1982, J.-B. Zuber and R. Stora (eds.) Amsterdam: North-Holland
1984, p. 561

\bibitem{FRT} Faddeev, L. D.,  Reshetikhin, N.Yu., Takhtajan, L.A.: Quantum
groups. In Braid groups, knot theory and statistical mechanics,
C.N. Yang and M.L. Ge (eds.) Singapore: World Scientific 1989

\bibitem{Fr} Frolov, S. A.: Gauge-invariant Hamiltonian formulation of lattice 
Yang-Mills theory and the Heisenberg double. 
Mod. Phys. Lett. {\bf A 10}, 2619 (1995) 

\bibitem{Gaudin} Gaudin, M.: La fonction d'onde de Bethe, Paris: Masson 1983

\bibitem{Miura}
Gardner, M., Greene, J., Kruskal, M., Miura, R.: Method for solving
the Korteveg-de Vries equation. Phys. Rev. Lett. {\bf 19}, 1095 (1967)

\bibitem{IK} Izergin, A. G., Korepin, V. E.: A lattice model related to
the nonlinear Schr\"odinger equation.
Soviet. Phys. Dokl. {\bf 26}, 653 (1981)

\bibitem{J} Jimbo, M.: A q--difference analogue of U(g) and the Yang--Baxter
equation. Lett. Math. Phys. {\bf 10}, 63 (1985)

\bibitem{JM} Jimbo M., Miwa T., Solitons and infinite dimensional Lie algebras.
Publ. RIMS, Kyoto University {\bf 19}, 943 (1983)

\bibitem{JSch} Jur\v co, B., Schlieker, M.: On Fock space
representations of quantized enveloping algebras related to non-commutative
differential geometry. J. Math. Phys. {\bf 36}, 3814 (1995)

\bibitem{JSch2} Jur\v co, B., Schlieker, M.: Quantized Lax equations and their
solutions. Preprint 1995, q-alg/9508001, to appear in Commun. Math. Phys.

\bibitem{BIK} Korepin V., Bogoliubov, N., Izergin, A. : Quantum inverse
scattering method and correlation functions, Cambridge: University Press 1993

\bibitem{K} Kostant, B.: The solution to a generalized Toda lattice and
representation theory. Adv. Math. {\bf 34}, 195 (1979)

\bibitem{KS} Kulish, P. P., Sklyanin, E. K., Integrable quantum field theories,
Lect. Notes Phys. {\bf 151}, 61 (1982)

\bibitem {Maillet} Maillet, J. M.: Lax equations and quantum groups, Phys.
Lett. {\bf B 245}, 480 (1990)

\bibitem{Resh} Reshetikhin, N.: Integrable discrete systems. In Quantum groups
and their applications in physics, Intl. School of Physics ``Enrico Fermi",
Varenna, 1994, L. Castellani and J. Wess (eds.), Bologna: Societ\`a Italiana di
Fisica 1995.

\bibitem{ReshSTS} Reshetikhin, N. Yu., Semenov-Tian-Shansky, M. A.:
Quantum R--matrices and factorization problems. J. Geom. Phys. {\bf
5}, 533 (1988)

\bibitem{ReySTS} Reyman, A. G., Semenov-Tian-Shansky, M. A.: Group theoretical
methods in the theory of finite-dimensional integrable systems. In Encyclopedia
of mathematical sciences, Dynamical systems VII,  Berlin, Heidelberg, New York:
Springer 1993

\bibitem{SWZ} Schupp, P., Watts, P., Zumino, B.: Bicovariant quantum algebras
and quantum Lie algebras. Commun. Math. Phys. {\bf 157}, 305 (1993)

\bibitem{SW} Segal, G., Wilson, G.: Loop groups and equations of KdV type.
Publ. Math. IHES {\bf 61}, 5 (1985)

\bibitem{STSDress}  Semenov Tian-Shansky, M. A.: Dressing transformation
and Poisson Lie group actions. Publ. RIMS, Kyoto University {\bf 21},
1237 (1985)

\bibitem{STS} Semenov Tian-Shansky, M. A.: Poisson Lie Groups, quantum duality
principle and the quantum double. Preprint 1993

\bibitem{Skly} Sklyanin, E. K.: Quantum variant of the method of the
inverse scattering. J. Soviet Math. {\bf 19}, 1546 (1982)

\bibitem{SoWa} Sogo, K., Wadati, M.: Quantum inverse scattering method and
Yang-Baxter relation for integrable spin system. Prog. Theor. Phys.
{\bf 68}, 85 (1982)

\bibitem{Symes} Symes, W.: System of Toda type, inverse spectral problems, and
representation theory. Inv. Math. {\bf 159}, 13 (1980)

\bibitem{Sweedler} Sweedler, M. E.: Hopf algebras, New York: Benjamin 1969

\bibitem{Thacker} Thacker, H. B.: Exact integrability in quantum field theory
and statistical systems. Rev. Mod. Phys. {\bf 53}, 253 (1981)

\bibitem{Zhang} Zhang, M. Q.: How to find the Lax pair from the quantum
Yang-Baxter equation. Commun. Math. Phys. {\bf 141}, 523 (1991)

\bibitem{Zum} Zumino, B., Introduction to the differential geometry of quantum
groups. In Math. Phys. X, Proc. X-th IAMP Conf. Leipzig, 1991,  K. Schm\"udgen
(ed.), Berlin: Springer 1992


\end{thebibliography}
\end{document}